\documentclass{PoS}
\usepackage{mathtools}

\title{Chiral and $U(1)_A$ restoration: Ward Identities and effective theories}

\ShortTitle{}

\author{\speaker{Angel G\'omez Nicola}
\\
        Departamento de F\'{\i}sica
Te\'orica and IPARCOS. Univ. Complutense. 28040 Madrid. Spain\\
        E-mail: \email{gomez@ucm.es}}

\author{Jacobo Ruiz de Elvira\\
        Albert Einstein Center for Fundamental Physics, Institute for Theoretical Physics,
University of Bern, Sidlerstrasse 5, CH--3012 Bern, Switzerland\\
        E-mail: \email{elvira@itp.unibe.ch}}

\author{Andrea Vioque-Rodr\'iguez\\
         Departamento de F\'{\i}sica
Te\'orica and IPARCOS. Univ. Complutense. 28040 Madrid. Spain\\
        E-mail: \email{avioque@ucm.es}}
        
        \author{Silvia Ferreres-Sol\'e\\
NIKHEF\\
Science Park 105,
NL-1098 XG,  Amsterdam
Netherlands
\\
        E-mail: \email{ferreres.sole@gmail.com}}

\abstract{We discuss our recent results regarding chiral and $U(1)_A$ restoration, both from the formal point of view of QCD Ward Identities (WI) and from an Effective Theory analysis provided by $U(3)$ Chiral Perturbation Theory (ChPT) at finite temperature.\ Our results lead to relevant conclusions regarding the behavior of chiral partners (in terms of susceptibilities) in the limit of exact restoration and provide useful results for lattice analysis. In addition, it helps to  understand the temperature dependence of lattice screening masses in terms of quark condensate combinations. The U(3) ChPT calculation supports the conclusions obtained within the WI analysis. Finally, the role of the thermal $f_0(500)$ state in chiral symmetry restoration, regarding   the scalar susceptibility, is also  discussed.}

\FullConference{XIII Quark Confinement and the Hadron Spectrum - Confinement2018\\
		31 July - 6 August 2018\\
		Maynooth University, Ireland}

\newcommand{\quarkcorl}{\langle {\cal T} (\bar \psi_l \psi_l (x)  \bar \psi_l \psi_l (0)\rangle}
\newcommand{\intT}{\int_0^\beta d\tau \int d^3 \vec{x}}

\newcommand{\condl}{\mean{\bar q q}_l}

\newcommand{\mean}[1]{\left\langle{#1}\right\rangle}

\newcommand{\im}{\mbox{Im}\,}
\newcommand{\Od}{{\cal O}}
\newcommand{\re}{\mbox{Re}\,}

\begin{document}

\section{Introduction}

Chiral and $U(1)_A$ symmetries, their nature and their possible restoration, are key elements of the QCD phase diagram. Chiral restoration is rather well understood from lattice simulations, 
which in the physical case, i.e., for $N_f=2+1$ flavors of masses $m_u=m_d=\hat m \ll m_s$, and for a vanishing baryon density support a crossover-like transition at a critical temperature of about $T_c\sim 155$ MeV~\cite{Aoki:2009sc,Borsanyi:2010bp,Bazavov:2011nk,Buchoff:2013nra,Bhattacharya:2014ara}. The chiral transition is customarily characterized by the inflection point of the light quark condensate $\condl$ and the maximum of the scalar susceptibility $\chi_S$~\cite{GomezNicola:2010tb}

\begin{eqnarray}
\condl(T)&=&\frac{\partial}{\partial \hat m}z(T),
\label{condef}\\
\chi_S (T)&=&-\frac{\partial}{\partial \hat m} \condl(T)=\int_T {d^4x \left[\quarkcorl-\condl^2(T)\right]},
\label{susdef}
\end{eqnarray}
where $\displaystyle \int_T dx\equiv \intT$ at finite temperature $T=1/\beta$, $\mean{\cdot}$ denote Euclidean finite-$T$ correlators and $z(T)=-\lim_{V\rightarrow\infty}(\beta V)^{-1}\log Z$ is the free energy density with $Z$ the QCD partition function.  As the system approaches the light chiral limit $\hat m/m_s\rightarrow 0^+$, $T_c$ decreases, the light quark condensate reduces, and the scalar susceptibility peak increases at $T_c$~\cite{Ejiri:2009ac},
 approaching  a second order phase transition expected for two massless flavors~\cite{Pisarski:1983ms,Smilga:1995qf}.

 The $U(1)_A$ symmetry can also be restored, although the nature of such restoration is not related to any spontaneous symmetry breaking, but to the presence of the chiral anomaly. Hence, $U(1)_A$ restoration takes place only asymptotically as the temperature increases, driven by the vanishing of the instanton density~\cite{Gross:1980br}. The possibility that $U(1)_A$ can be restored at a temperature close to the chiral transition has profound implications regarding its universality class, the oder of the transition~\cite{Pisarski:1983ms,Pelissetto:2013hqa}  and the behavior near the critical end point at finite temperature and baryon chemical potential~\cite{Mitter:2013fxa}. This would also directly affect the way in which different hadron states degenerate near the transition (chiral partners). Considering in particular the $0^{++}$ and pseudoscalar $0^{-+}$  meson nonets; if the chiral group $SU_L(2)\times SU_R(2)$ is restored, the pion is expected to degenerate with the  $\sigma_l$, the light component of the $\sigma/f_0(500)$~\cite{Hatsuda:1985eb,Bernard:1987im,Krippa:2000jh} and so would do the $a_0(980)$ with the $\eta_l$, the light component of the $\eta/\eta'$ pair. If $U(1)_A$ is restored, the degeneration pattern would be $\pi-a_0$ and $\sigma_l-\eta_l$, i.e., octet members with same quantum numbers but opposite parity. In terms of quark bilinears,
 
\begin{eqnarray}
\pi^a&=&i\bar\psi_l\gamma_5\tau^a\psi_l=P^a (a=1,2,3),\quad 
\delta^a=\bar\psi_l \tau^a \psi_l=S^a (a=1,2,3),\nonumber\\
\eta_l&=&i\bar\psi_l \gamma_5 \psi_l,\hspace*{3.4cm}
\sigma_l=\bar\psi_l \psi_l,
\label{bilinears}
\end{eqnarray}
 where $\delta^a$ correspond to the $a_0(980)$. The rest of the octet members will satisfy also similar degeneration patterns. Namely,  the $K(700)$ (or $\kappa$) versus the kaon for $I=1/2$, and the $f_0(980)-f_0(500)$ pair versus the $\eta-\eta'$ for the $I=0$ octet and singlet members.  Actually, the restoration of the $U(1)_A$ symmetry also affects the temperature dependence of the $\eta-\eta'$ mixing, which is expected to approach the so called ideal mixing as the temperature increases~\cite{Lenaghan:2000ey,Ishii:2016dln,Rennecke:2016tkm}, i.e., the $\eta$ and $\eta'$  become states with pure light and strange quark content, respectively. 

Nevertheless, there is still not full agreement among lattice collaborations on whether the $U(1)_A$ symmetry is restored close enough to the chiral $O(4)$ one. On the one hand, for $N_f=2+1$ flavors and physical quark masses, the analysis of~\cite{Buchoff:2013nra,Bhattacharya:2014ara} shows degeneracy of $U(1)_A$ partners well above the $O(4)$ ones. On the other hand, $N_f=2$ analyses in the chiral limit~\cite{Aoki:2012yj,Cossu:2013uua,Tomiya:2016jwr}  and in the massive case~\cite{Brandt:2016daq}, indicate  $U(1)_A$ restoration very close above the chiral one. 
 
Our theoretical approach is based on the use of Ward Identities (WI) connecting the bilinears defined in~\eqref{bilinears}, both formally in QCD and within the low-energy hadronic description provided by Chiral Perturbation Theory, including the $\eta'$ anomalous sector in the $U(3)$ formalism~\cite{Nicola:2013vma,Nicola:2016jlj,GomezNicola:2017bhm,Nicola:2018vug}. Such analysis has been completed for the full scalar and pseudoscalar nonets and allows one to relate susceptibilities with combinations of quark condensates and differences of  partner susceptibilities with physical vertices. In particular, as we will show in section \ref{WI}, the symmetry transformation properties of such identities lead to interesting consequences regarding the relation between chiral and $U(1)_A$ restoration. In addition, they allow one to explain quite accurately the scaling with temperature of lattice screening masses.  
 
 A closely related analysis, which we  also review here, is the study of the role of the $f_0(500)$ state in chiral symmetry restoration~\cite{Nicola:2013vma,Ferreres-Sole:2018djq}. In particular, we show that the scalar susceptibility $\chi_S(T)$ saturated by the pole of the $f_0(500)$ at finite temperature describes remarkably well the expected crossover behavior around the transition, in agreement with lattice data. We will give more details of this approach in section \ref{sus}.

 \section{Ward Identities: chiral vs $U(1)_A$ restoration and screening masses}
 \label{WI}
 
 As stated above, the use of certain WI sheds light on the relation between chiral  and $U(1)_A$ restoration. The following identities are particularly useful in that respect~\cite{Nicola:2016jlj,GomezNicola:2017bhm,Nicola:2018vug}:
 
  \begin{align}
&\chi_P^\pi(T)=-\frac{\condl(T)}{\hat m},
\label{wichip}\\ 
&\chi_P^{ll}(T)=-\frac{\condl(T)}{\hat m}-\frac{4}{\hat m^2} \chi_{top},
\label{chipetal}
\end{align}
where $\chi_P^\pi$ and $\chi_P^{ll}$ are respectively the pseudoscalar susceptibilities (zero momentum correlators) associated to the $\pi$ and $\eta_l$ bilinears in~\eqref{bilinears}, while 

\begin{equation}
\chi_{top}(T)\equiv -\frac{1}{36}\chi_P^{AA}(T)=-\frac{1}{36}\int_T dx  \langle \mathcal{T} A(x) A(0) \rangle,
\label{chitopdef}
\end{equation} 
with 
  $A(x)=\frac{3g^2}{16\pi^2}\mbox{Tr}_c G_{\mu\nu}\tilde G^{\mu\nu}$,
is the topological susceptibility. 

The combination of~\eqref{wichip} and~\eqref{chipetal} plus an additional identity for the crossed pseudoscalar susceptibility $\chi_P^{ls}$ between $\eta_l$ and $\eta_s=i\bar s \gamma_5 s$, allows one to write:

\begin{equation}
\chi_P^{ls}(T)=-2\frac{\hat m}{m_s} \chi_{5,disc}(T)=-\frac{2}{\hat m m_s}\chi_{top}(T),
\label{wils5}
\end{equation}
where 
\begin{equation}
\chi_{5,disc}(T)=\frac{1}{4}\left[\chi_P^\pi(T)-\chi_P^{ll}(T)\right]
\label{chi5def}
\end{equation}
is the order parameter used in lattice simulations to study $O(4)\times U(1)_A$ restoration, according to our previous discussion on partner degeneration. 

The importance of~\eqref{wils5} is the following: under an axial $SU(2)_A$ transformation $\psi_l\rightarrow e^{i\gamma_5 \alpha_a\tau^a/2}\psi_l$ we have
\begin{equation}
\eta_l(x)\rightarrow i\bar\psi_l(x)\gamma_5 e^{i\gamma_5 \alpha_a\tau^a}\psi_l(x)=i\bar\psi_l(x)\gamma_5 \cos (\alpha_a\tau^a)\psi_l(x)-\bar\psi_l(x) \sin (\alpha_a\tau^a)\psi_l(x),
\label{chilsvanishing1}
\end{equation}
with $a=1,2,3$. Thus, for the  particular choice 
\begin{equation}
\alpha_b=\pi/2 \quad{\rm and}\quad \alpha_{a\neq b}=0, \qquad 
\label{trans}
\end{equation}
 we have
   \begin{equation}
\eta_l(x)\rightarrow -\bar\psi_l(x) \tau^b \psi_l(x)=-\delta^b (x)\Rightarrow P_{ls}(x)\rightarrow -\mean {{\cal T}  \delta^b (x) \eta_s(0)} =0,
\label{chilsvanishing2}
\end{equation}
where $P_{ls}$ is the $\eta_l\eta_s$ coorrelator and we have used that $\eta_s$  is invariant under $SU(2)_A$ transformations and the last correlator vanishes by parity. 
Therefore, if $O(4)$ is completely restored so that the correlators related by $SU(2)_A$ transformations degenerate, $\chi_P^{ls}$ should vanish.
Consequently, the relation~\eqref{wils5} together with the previous argument leads to the conclusion that in the phase where $\delta-\eta_l$ degenerate ($O(4)$) $\chi_{5,disc}$ should vanish and $\pi^a-\eta$ degenerate as well ($O(4)\times U(1)_A$). The same identity implies also the vanishing of $\chi_{top}$. This argument favors then a $O(4)\times U(1)_A$ pattern, at least from the formal viewpoint, along the lattice results in~\cite{Aoki:2012yj,Cossu:2013uua,Tomiya:2016jwr,Brandt:2016daq}.
In the physical case one finds larger uncertainties for $O(4)$ $\delta-\eta_l$ degeneration~\cite{Buchoff:2013nra,Bhattacharya:2014ara}, which together with the strangeness contribution might lead to a larger gap between those transitions~\cite{GomezNicola:2017bhm,Nicola:2018vug}. 

\begin{figure}
\centerline{\includegraphics[width=9cm]{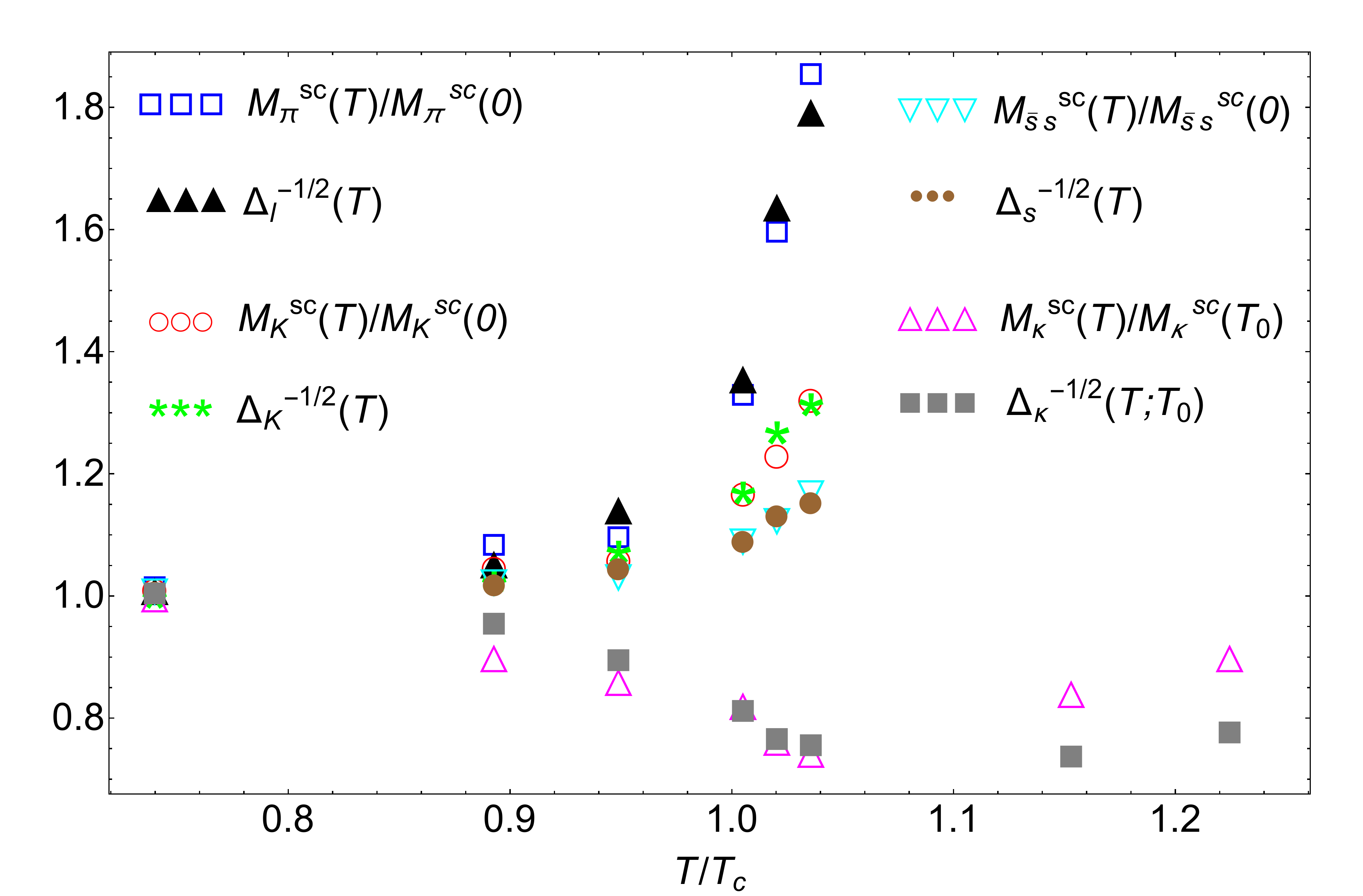}}
\caption{Comparison of pseudoscalar screening mass ratios and subtracted condensates for the four channels $\pi$, $K$, $\bar s s$ and $\kappa$.  The  lattice data are taken from 
~\cite{Cheng:2007jq} (condensates) and~\cite{Cheng:2010fe}  (masses).}
\label{fig:corrfour}
\end{figure}

Another implication of the WI discussed above is that they allow one to understand the temperature dependence of lattice meson screening masses~\cite{Nicola:2016jlj,Nicola:2018vug}; since the susceptibilities $\chi_i$ correspond to zero momentum correlators, one can assume for meson states a scaling of the form $M_i(T)/M_i(0)\sim \left[\chi_i (T)/\chi_i(0)\right]^{-1/2}$ and use for $\chi_i(T)$ the WI relating them to quark condensate combinations. The latter have to be properly subtracted to avoid lattice divergences. In Fig.\ref{fig:corrfour} we show such comparison of scalings predicted by the WI for lattice data of the same collaboration with the same lattice action and resolution. The $\Delta_i$ correspond to subtracted condensates defined in terms of two fit parameters (see~\cite{Nicola:2016jlj,Nicola:2018vug} for details). Data above $1.1 T_c$ are not fitted. The agreement is remarkably good and the WI also explain the strength of the temperature growth of the different channels. For instance, the pion channel would grow like $\condl^{-1/2}$ according to~\eqref{wichip}, while in the $K$ and $\bar s s$ channels there is a $\langle \bar s s\rangle$ condensate contribution softening the temperature behavior.

\section{$U(3)$ Chiral Perturbation Theory analysis of chiral and $U(1)_A$ restoration}
\label{chpt}  
  
The discussion in section \ref{WI} has dealt with formal WI derived from QCD. A particular hadronic low-energy realization of those results is provided by $U(3)$ Chiral Perturbation Theory (ChPT), which is the most general framework describing the $\pi$, $K$, $\eta$ and $\eta'$ states. In order to incorporate properly the large mass of the singlet $\eta_0$ due to the axial anomaly, the $U(3)$ ChPT framework relies on the large-$N_c$ regime ~\cite{Witten:1979vv,HerreraSiklody:1996pm,Kaiser:2000gs,Guo:2012ym}, so that the chiral counting is extended to include  $1/N_c$  in a general parameter $\delta$ such that $M^2, E^2, T^2,\hat m,m_s =\Od(\delta)$ and $1/N_c=\Od(\delta)$, where $M,E,T$ are typical meson masses, energies and temperatures.  

\begin{figure}
\centerline{\includegraphics[width=8cm]{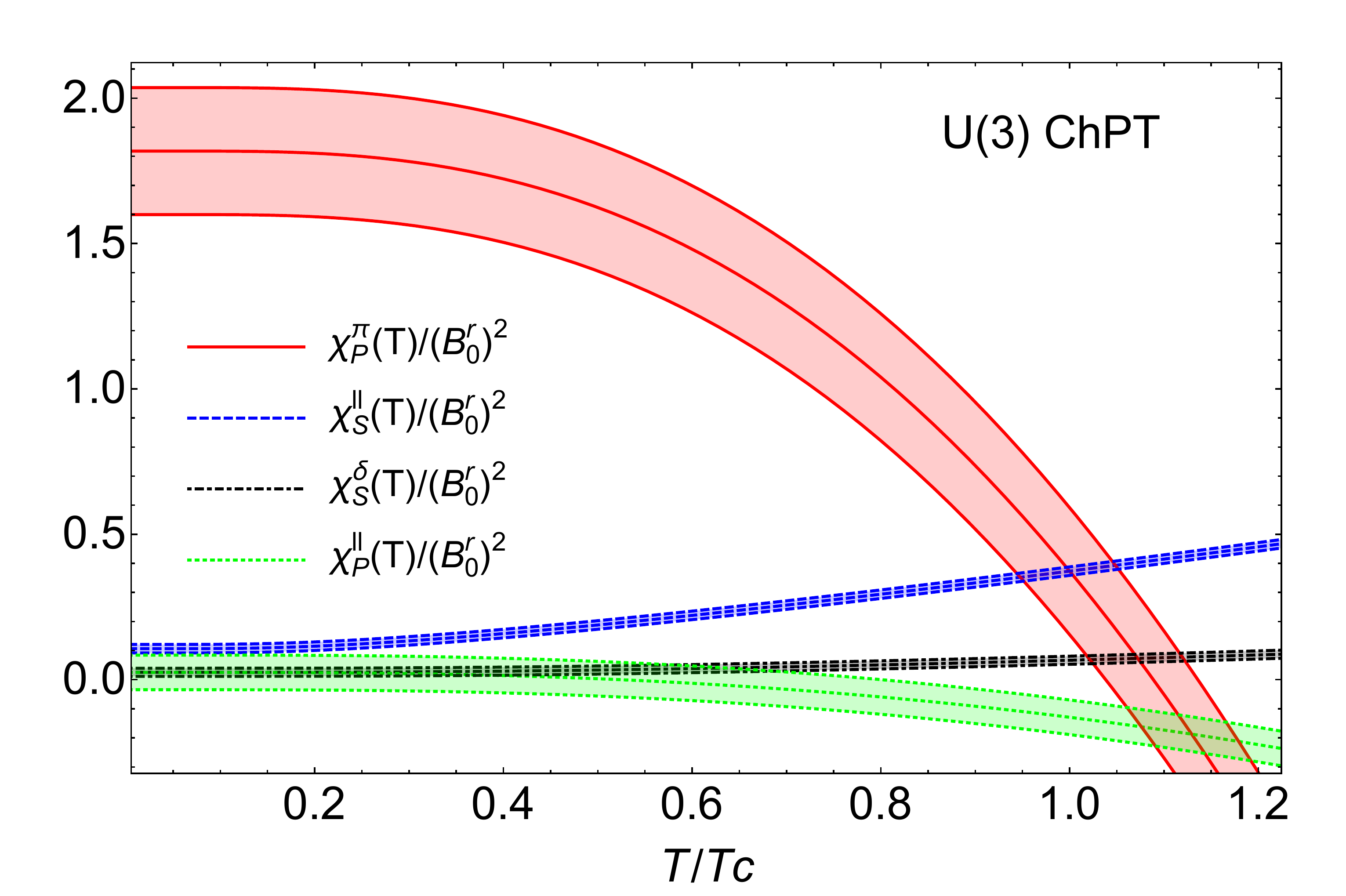}\includegraphics[width=8cm]{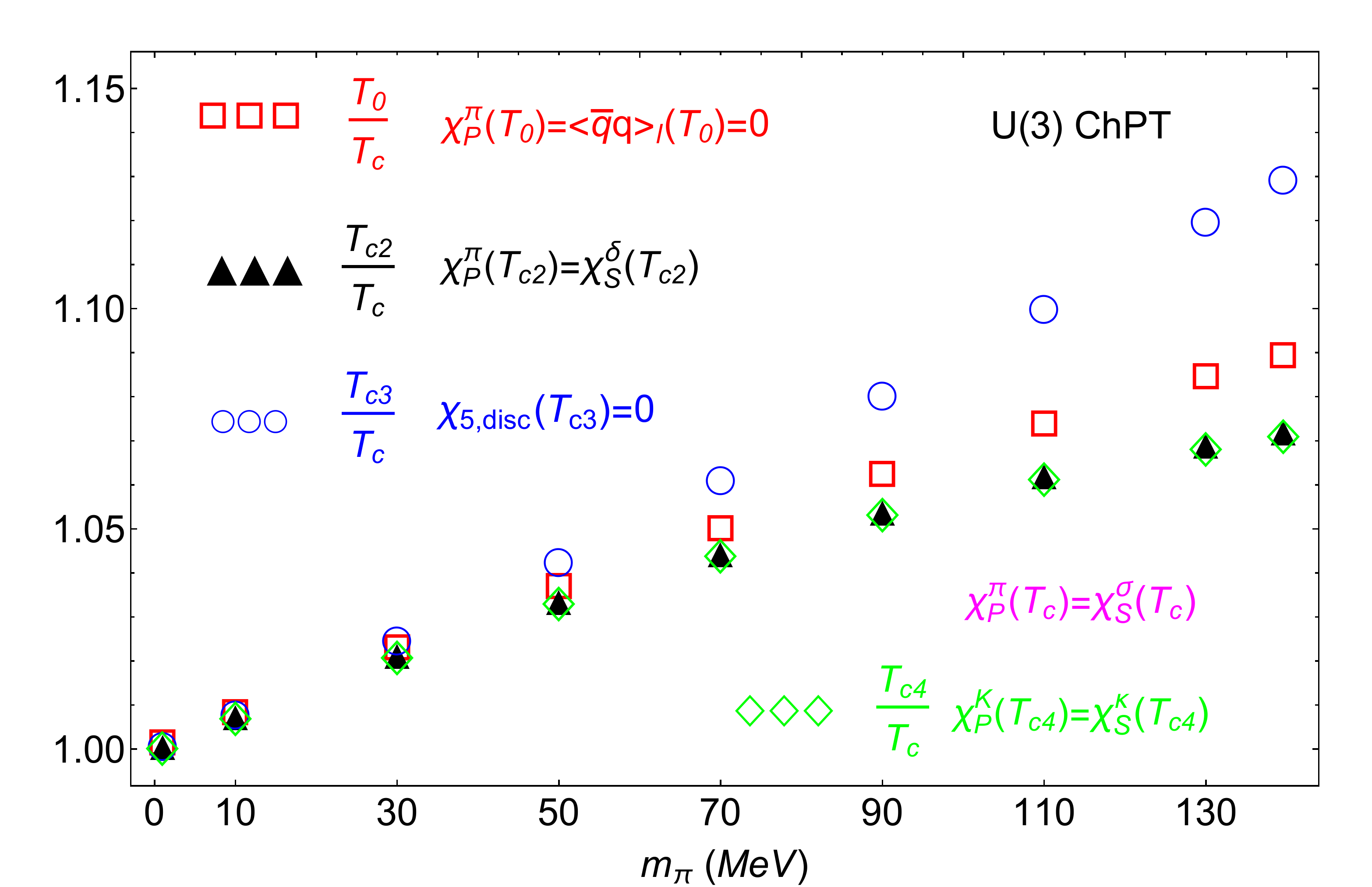}}
\caption{Left:$U(3)$ ChPT results for the isospin $I=0,1$ $\pi,\sigma,\eta_l,\delta$ susceptibilities. Right: Evolution towards the chiral limit of the different $O(4)$ and $U(1)_A$ temperatures}
\label{fig:u3chpt}
\end{figure}

Within that framework, we have analyzed in~\cite{Nicola:2018vug} the different susceptibilities involved in the chiral and $U(1)_A$ degeneration of the scalar and pseudoscalar nonets at finite temperature. Apart from checking the WI in this effective theory realization, we confirm the restoration pattern discussed in section \ref{WI}. In Fig.\ref{fig:u3chpt} we show our results for the susceptibilities corresponding to the four bilinears in~\eqref{bilinears}. They confirm that the chiral $O(4)$ and $U(1)_A$ symmetries remain close in terms of partner degeneration in the physical massive case, with the $U(1)_A$ degeneration of $\pi-\delta$ taking place at a temperature around $1.07 T_c$ with $T_c$, the temperature where $\chi_S^{ll}$ (corresponding to the $\sigma$ state) and $\chi_P^\pi$ match.  The bands in that figure correspond to the numerical uncertainties of the Low Energy Constants (LEC) involved.

  Furthermore, in the same figure we also show the trend towards the chiral limit of the different degeneration temperatures for  the nonet members. It can be clearly seen that all tend to the same value as the chiral limit is approached. Since in the chiral limit $O(4)$ restoration is meant to be exact, these results confirm the conclusions obtained in section~\ref{WI}. In addition, we have obtained in~\cite{Nicola:2018vug}  that the temperature dependence of $\chi_{5,disc}(T)$ in $U(3)$ ChPT is the same as that of $\condl (T)$ near the chiral limit, and they become close in the massive case. Once again, this confirms that these two order parameters and their corresponding restoration transitions are linked,  consistently with our analysis in terms of WI.  Finally, within the $U(3)$ framework, we have also obtained  that the $\eta-\eta'$ mixing angle approaches the ideal limit around the critical region~\cite{Nicola:2018vug}.

 \section{Describing the scalar susceptibility by the thermal $f_0(500)$ pole}
 \label{sus}
 
\begin{figure}
\centerline{\includegraphics[width=9cm]{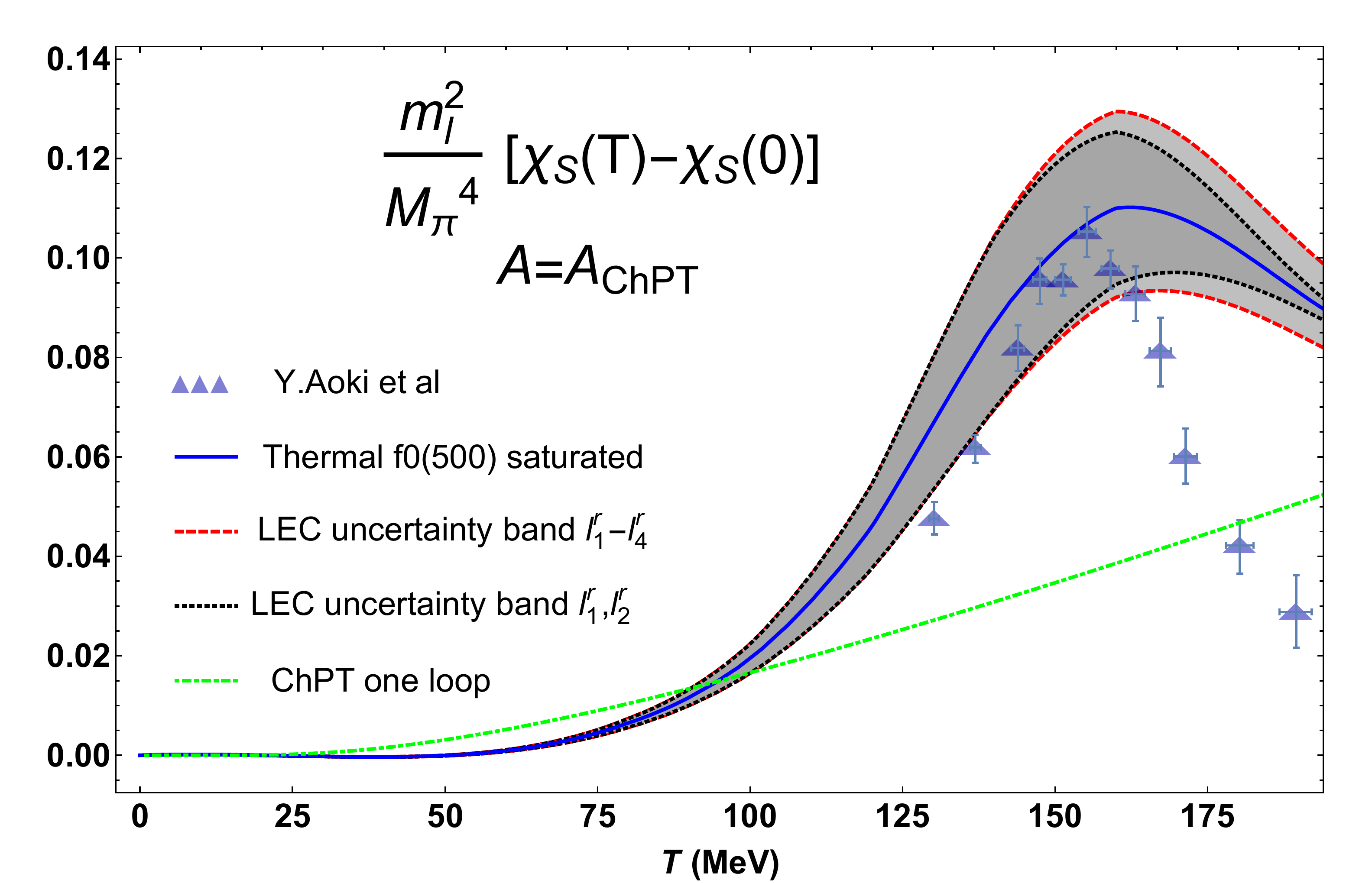}}
\caption{Saturated scalar susceptibility  including the uncertainties coming from the LEC involved. The lattice data and errors are from~\cite{Aoki:2009sc}.}
\label{fig:unitsuslec}
\end{figure}

 Another recent important line of research concerning chiral restoration has been the analysis of the role of the thermal $f_0(500)$ to describe the scalar susceptibility~\cite{Nicola:2013vma,Ferreres-Sole:2018djq}.  One can show that under certain assumptions, the scalar susceptibility~\eqref{susdef} can be related to the zero momentum propagator of the $\sigma/f_0(500)$ state. On the other hand, Unitarized ChPT provides a reliable framework to describe the $f_0(500)$  as a resonance in the second Riemann sheet (2RS) of the $\pi\pi$  scattering amplitude~\cite{Pelaez:2015qba}, including finite temperature effects, for instance through the so called Inverse Amplitude Method (IAM) ~\cite{Dobado:2002xf}. In that framework, the $I=J=0$ partial wave reads
 
 \begin{equation}
t_{IAM}(s;T)=\frac{t_2(s)^2}{t_2(s)-t_4(s,T)}.
\label{iam}
\end{equation}
 where $t_2+t_4+\dots$ correspond to the standard ChPT series and $t_{IAM}$ satisfies 
 $$\im t_{IAM}(s;T)=\sigma_T(s)\vert t_{IAM}(s;T)\vert^2$$  with $\sigma_T(s)=\sqrt{1-\frac{4M_\pi^2}{s}}\left[1+2n_B(\sqrt{s}/2)\right]$ and  $n_B(x)=(1-e^{x/T})^{-1}$ the Bose-Einstein distribution function. The amplitude defined through~\eqref{iam} develops a pole in the 2RS which corresponds to the $f_0(500)$ at finite temperature. Around the pole, 
 
 \begin{equation}
t^{II}=\frac{1}{16\pi}\frac{g_{\sigma\pi\pi}^2}{s-s_p}+\dots
\label{unitpoleexp}
\end{equation}
with $s_p(T)=(M_p(T)-i\Gamma_p(T)/2)^2$ and $g_{\sigma\pi\pi}$ the effective $\sigma\pi\pi$ effective coupling. Regarding~\eqref{unitpoleexp}  as the exchange of the $f_0$ state with self-energy $\Sigma (s_p)=s_p$, taking into account that $\im\Sigma(0)=0$ and assuming that the result is not affected much by the variation of $\re\Sigma$ from $s=0$ to $s=s_p$, one has for the unitarized scalar susceptibility saturated by the thermal $f_0(500)$,

  \begin{equation}
\chi_S^U(T)=A\frac{M_\pi^4}{4m_l^2}\frac{M_S^2(0)}{M_S^2(T)},
\label{susunit}
\end{equation}
with 
 \begin{equation}
M_S^2(T)=\re s_p (T)=M_p^2(T)-\frac{1}{4}\Gamma_p^2(T),
\label{scalarmass}
\end{equation}
 and $A$ a proper normalization constant which accounts partially for the uncertainties in this approach. Choosing $A$ to match the perturbative ChPT one-loop result for $\chi_S$ at $T=0$ $A_{ChPT}\simeq 0.15$,~\eqref{susunit} provides a very good description of lattice data, as we show in Fig. \ref{fig:unitsuslec}, taken from~\cite{Ferreres-Sole:2018djq}. The bands correspond to the uncertainty provided by the LEC involved, which in this case are those related to ChPT pion scattering, namely the renormalized $l_1^r,l_2^r,l_3^r,l_4^r$ in~\cite{Gasser:1983yg}. The results are mostly sensitive to $l_1^r$ and $l_2^r$, as the figure shows, $l_3^r$ and $l_4^r$ coming only from the renormalization of $M_\pi$ and $F_\pi$. Their central values and  uncertainties are taken from~\cite{Hanhart:2008mx}, where the LEC are fitted to experimental data and give a good agreement with the PDG for the $f_0(500)$ and $\rho(770)$ $T=0$  poles.

\begin{figure}
\centerline{\includegraphics[width=8cm]{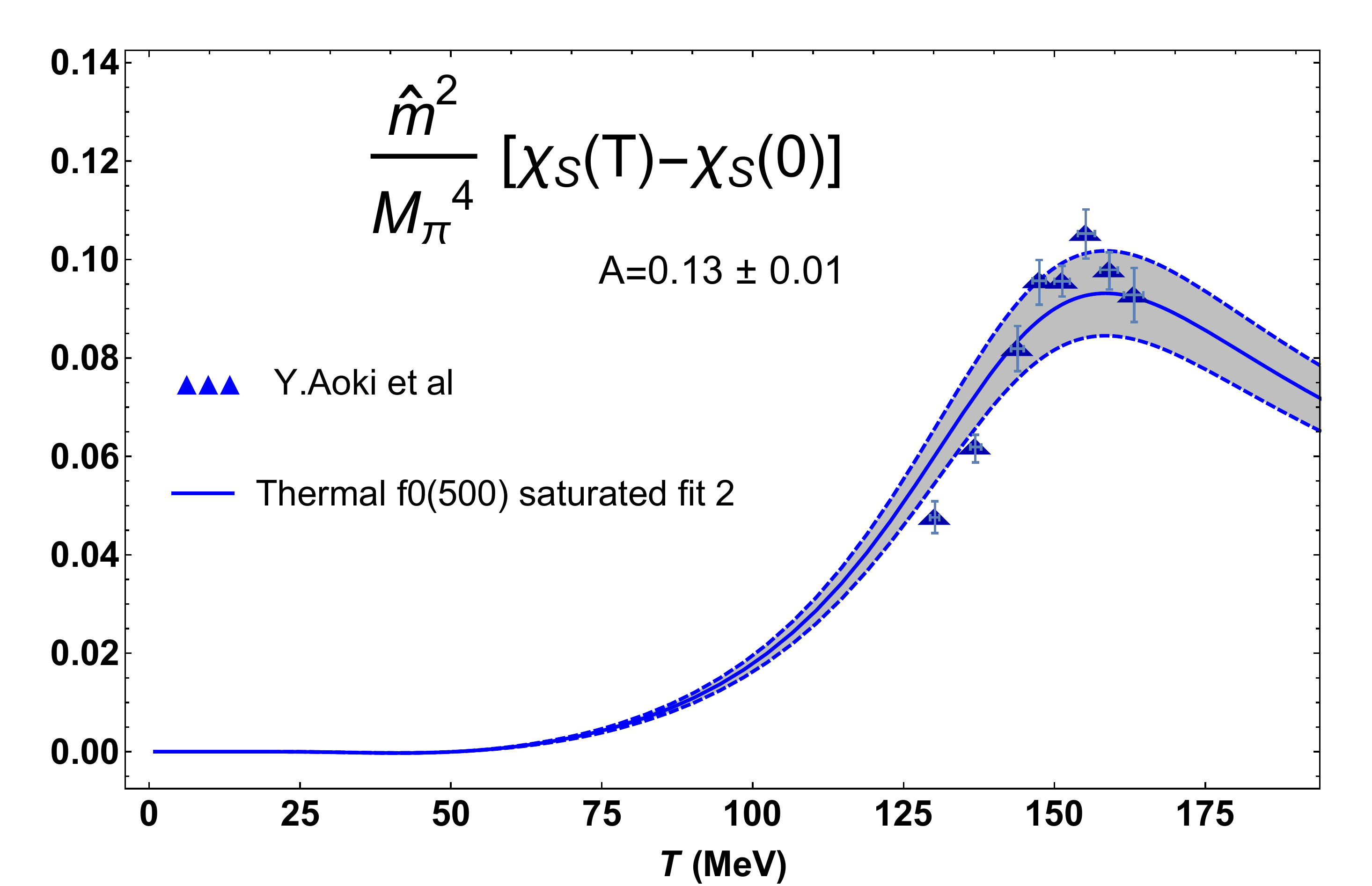}\includegraphics[width=8cm]{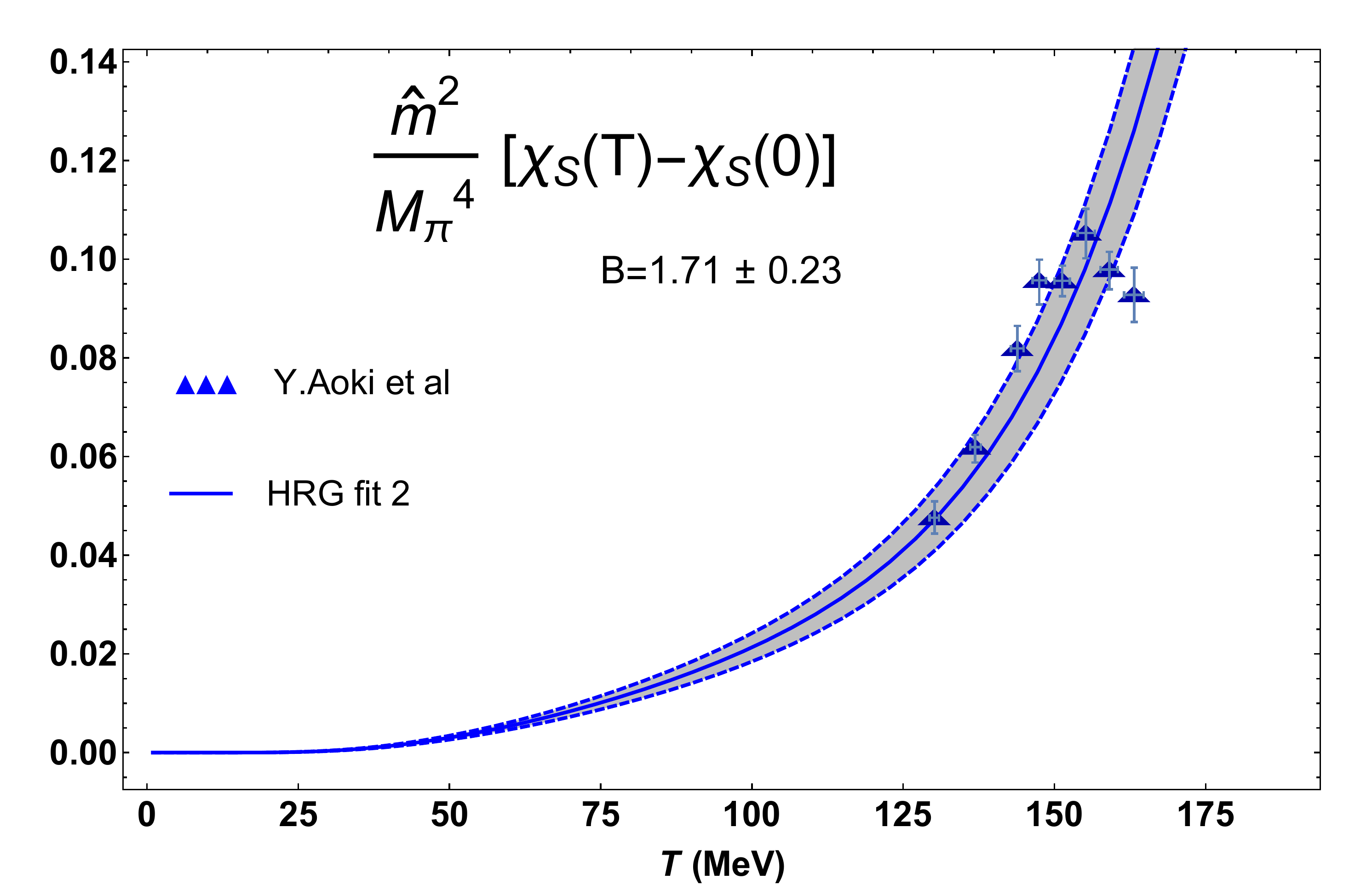}}
\caption{Comparison of fits to lattice data for the thermal $f_0(500)$ approach (left) and the HRG (right). The quoted uncertainties in the fit parameters and the bands correspond to the 95\% confidence level. The lattice data and errors are from~\cite{Aoki:2009sc}. }
\label{fig:fits}
\end{figure}

 The previous result shows that the unitarized susceptibility~\eqref{susunit} can describe lattice data within the $T=0$ uncertainties. A more quantitative evaluation of the  predictive power of this approach can be obtained by comparing it with the well established method of the Hadron Resonance Gas (HRG), as carried out in~\cite{Ferreres-Sole:2018djq}. In particular, to derive the scalar susceptibility,  a HRG approach has been used where the mass dependence of the different hadrons is obtained through a constituent NJL-like approach as described in~\cite{Leupold:2006ih,Jankowski:2012ms}. In order to compare both approaches,   $A$  in~\eqref{susunit} has been left as a fit parameter,  setting also  a normalization parameter $B$ for the HRG free energy, and fitting both to lattice data. In Fig.\ref{fig:fits} we show  the results for such a fit as given in~\cite{Ferreres-Sole:2018djq}, including temperature values up to 163 MeV. The HRG tends to give a better description, as expected, for data below the maximum, but the saturated approach  can account better for the data around the transition peak. For the fit shown in the figure, the $\chi^2/$dof equals  4.9 for the $f_0(500)$ fit and 10.3 for the HRG one. For comparison, taking out the lattice points above the peak   reduces the HRG $\chi^2/$dof to 1.3, increasing the saturated one to 6.2.

 \section{Conclusions}
 
 We have shown that the use of formal QCD Ward Identities allows one to extract powerful conclusions regarding chiral $O(4)$ and $U(1)_A$ restoration. In particular, formal $O(4)$ restoration  in terms of the $\eta_l$ and $\delta^a$ ($a_0(980)$ meson) partners points to a $O(4)\times U(1)_A$ pattern in terms of $\pi^a$ and $\eta_l$ degeneration and the vanishing of the topological susceptibility. The WI analysis also allows one to determine the temperature scaling of lattice meson screening masses in terms of quark condensates, which fits well with lattice data and helps to understand their qualitative   behavior.  We have also shown the results of a $U(3)$ calculation at finite temperature for the susceptibilities of the scalar and pseudoscalar nonets, which confirms the findings of the WI analysis. In particular, in the chiral limit we find that  the $O(4)$ and  $O(4)\times U(1)_A$ partner degeneration temperatures become identical and the order parameter $\chi_{5,disc}$ scales like the light quark condensate. 
 
 The result of a recent analysis of the role of the thermal $f_0(500)$ to describe the scalar susceptibility has also been reviewed here. Assuming that the scalar susceptibility is saturated by the $f_0(500)$ resonance at finite temperature, whose pole is calculated from a unitarized ChPT approach, provides a very accurate description of lattice data, which improves over the standard Hadron Resonance Gas near the transition peak.

 \section*{Acknowledgments}
Work partially supported by  research contract FPA2016-75654-C2-2-P  (spanish ``Ministerio de Econom\'{\i}a y Competitividad") and by the Swiss National Science Foundation.

\end{document}